\documentclass[aps,prd,amssymb,amsmath,nobibnotes,nofootinbib,superscriptaddress, reprint, 12pt ]{revtex4}
\usepackage{amsfonts,amsmath,hyperref,url, color}
\allowdisplaybreaks
\usepackage{bm, bbm}
\usepackage{graphicx}
\usepackage{mathtools}
\usepackage{t1enc}
\usepackage{hepnames}
\usepackage{tensor}
\usepackage{comment}
\usepackage{xcolor}
\usepackage{appendix}

\usepackage{dcolumn}

\usepackage{adjustbox,booktabs}

\usepackage{makecell}
\usepackage{array}

\newcommand{\bc}{\begin{center}}
\newcommand{\ec}{\end{center}}
\newcommand{\be}{\begin{eqnarray}}
\newcommand{\ee}{\end{eqnarray}}
\newcommand{\bs}{\begin{slide}}
\newcommand{\es}{\end{slide}}
\newcommand{\la}{\langle}
\newcommand{\ra}{\rangle}
\newcommand{\bi}{\begin{itemize}}
\newcommand{\ei}{\end{itemize}}
\newcommand{\nn}{\nonumber}

\begin{document}

\title{Deformation-induced CPT violation \\ in entangled pairs of neutral kaons}

\author{Andrea Bevilacqua}
\email{andrea.bevilacqua@ncbj.gov.pl}
\affiliation{National Centre for Nuclear Research, ul. Pasteura 7, 02-093 Warsaw, Poland}
\author{Jerzy Kowalski-Glikman}
\email{jerzy.kowalski-glikman@uwr.edu.pl}
\affiliation{University of Wroc\l{}aw, Faculty of Physics and Astronomy, pl.\ M.\ Borna 9, 50-204
Wroc\l{}aw, Poland}
\affiliation{National Centre for Nuclear Research, ul. Pasteura 7, 02-093 Warsaw, Poland}
\author{Wojciech Wi\'slicki}
\email{wojciech.wislicki@ncbj.gov.pl, corresponding author}
\affiliation{National Centre for Nuclear Research, ul. Pasteura 7, 02-093 Warsaw, Poland}

\date{\today}

\begin{abstract}
    In this paper we consider description of kaon -- anti-kaon interference in the context of a theory with deformed $\cal CPT$ symmetry. In the case of such theoretical models, deviations from the standard $\cal CPT$ invariance is related to the momentum carried by the particles; in particular the rest masses of particles and antiparticles are equal. We find that the decay intensity of kaon -- anti-kaon pair  has three contributing terms: the correct-parity, the wrong-parity, and the interference between them, all of which are affected by deformation. Using the fact that the presence of such terms were not observed we estimate the magnitude of deformation parameter $\kappa \gtrsim 10^{18}$ GeV, very close to the expected Planck mass scale. This raises hopes that the effect, if exists, could be detected in a near future accelerator experiments. An uncertainty of the energy measurement is crucial for accuracy of these predictions.
\end{abstract}
\maketitle

\section{Introduction}

It was realized some time ago \cite{Amelino-Camelia:1999hpv} that, contrary to what was believed at the time, quantum gravity phenomenology is not only feasible, but that some effects of quantum gravitational origin can be observed using current or near-future technologies. Today, with the advent of multi-messenger astronomy, new observational opportunities using astrophysical sources of high energy particles are opening up (see \cite{Amelino-Camelia:2008aez}, \cite{Addazi:2021xuf}, \cite{AlvesBatista:2023wqm} for  recent reviews). Unfortunately, there are still obstacles to the implementation of this research program in the astrophysical realm, such as low event statistics and poor understanding of the astrophysical sources and the medium through which the cosmic messengers travel. It is therefore tempting to look for opportunities to observe quantum gravity-induced effects in the realm of better controlled and offering much better statistics terrestrial high-energy experiments.

One of the most promising theoretical frameworks predicting possible observable effects at energies much lower than the characteristic scale of quantum gravity, the Planck energy $\kappa\simeq 10^{19}$ GeV, is the quantum gravity induced Lorentz Invariance Violation (LIV)\footnote{This general term describes both the quantum gravity induced manifest (spontaneous) Lorentz invariance breaking, as in the Kostelecky model \cite{Kostelecky:1988zi}, and theories, in which Lorentz invariance is deformed, like in Doubly Special Relativity \cite{Amelino-Camelia:2000stu} or Relative Locality \cite{Amelino-Camelia:2011lvm}. }. For example, a large class of LIV models predict the energy dependence of the velocity of massless particles, an effect which might be observable in principle \cite{Amelino-Camelia:1997ieq}, \cite{Addazi:2021xuf}, \cite{AlvesBatista:2023wqm}.

Another potentially observable effect arising as a consequence of quantum gravity might be possible deviations from $\cal CPT$ symmetry. According to Greenberg theorem \cite{Greenberg:2002uu} violation of $\cal CPT$ symmetry is a consequence of Lorentz symmetry breaking and vice versa. Specifically, models of violation of the $\cal CPT$ and Lorentz symmetry invariance were proposed in the context of extensions of the Standard Model with electroweak interactions coupled to gravity \cite{kostelecky_1}, or models of gravity with decoherence \cite{bernabeu_1} at energy scale of the Planck mass $m_P\simeq 10^{19}$ GeV.

In a recent development, deformations of $\cal CPT$ symmetry have been studied in the context of $\kappa$ deformations \cite{Arzano:2016egk}, \cite{Arzano:2019toz}, \cite{Arzano:2020rzu}, \cite{Arzano:2020jro}, \cite{bevilacqua_1}. In this case we have to do with modification of both continuous (Poincar\'e) and discrete symmetries, but all of the symmetries are still present and are not violated. The distinctive feature of the models with deformed (as opposite to violated) $\cal CPT$ symmetry is that the deviations from the standard $\cal CPT$-invariance are momentum dependent. In particular, the rest masses of particles and antiparticles are equal. As a consequence the decay rates of particles and antiparticles at rest are equal, but they start differ for the moving ones \cite{Arzano:2019toz}, \cite{Lobo:2020qoa}.
Although the size of this mismatch is tiny at energies available so far in accelerators, and is still beyond reach of the accuracy of detectors built at present, its quadratic dependence on particle momenta, as well as rapidly developing acceleration and detection technologies, motivate theoretical efforts in this direction. 

A natural experimental framework where such a hypothesis can be examined is provided by any particle-antiparticle system with well-controlled momenta of both particles. When a particle-antiparticle pair is produced in an unstable resonance decay, their total linear momentum vanishes in the resonance rest frame. 
The decay products are entangled, and quantum interference provides a sensitive tool to compare lifetimes by measuring their decay and interference spectrum in their relative decay times.
With this possibility in mind, the authors of  \cite{bernabeu_2} proposed a parametrization of the entangled state of two neutral kaons in the case when $\cal CPT$-symmetry is violated. In the present paper we reconsider the production of an entangled pair of neutral kaons in the context of $\kappa$-$\cal CPT$ deformation proposed in \cite{Arzano:2019toz}, \cite{Arzano:2020jro}, \cite{bevilacqua_1}.

In the following Sect.\ II we will recall briefly the theory of entangled pairs of neutral kaons and its modification proposed in \cite{bernabeu_2}. Then in Sect.\ III we discuss how the behavior of kaons is modified in deformed theory. Finally, Sect.\ IV is devoted to a discussion of bounds on the deformation parameter $\kappa$ that arise if no deformation effects are observed.

\section{Entangled pairs of neutral kaons}

If one considers the $K^0\overline K^0$ pair in the non-deformed context, created in a decay of the $\Phi^0 (1020)$ resonance with $J^{PC}=1^{--}$ and flying apart with momenta $\pm \mathbf p$ in the centre-of-mass frame of the whole system, requirements of the Bose-Einstein statistics and conservation of the total angular momentum provide their state to be
\be
\label{eq1}
|\psi\ra & = & \frac{1}{\sqrt 2}\left(|K^0(\mathbf p)\ra|\overline K^0(-\mathbf p)\ra -|\overline K^0(\mathbf p)\ra|K^0(-\mathbf p)\ra\right). 
\ee
Expressing the state $|\psi\ra$ in the basis of the neutral kaon mass eigenstates $|K_{S,L}\ra$
\be
\label{eq3}
|K_{S,L}\ra & = & \frac{1}{\sqrt{2(1+|\varepsilon\pm\delta|^2)}}\left[(1+\varepsilon\pm\delta)\big|K^0\ra \pm (1-\varepsilon\mp\delta)|\overline K^0\ra\right] 
\ee
one makes use of two experimentally measurable complex parameters quantifying the $\cal CP$ and $\cal CPT$ violation\footnote{We use the notation of Ref. \cite{pdg}}.
In Eq. (\ref{eq3}) the upper signs refer to $K_S$ and the lower ones to $K_L$.
The experimental value of $|\varepsilon|$ is of order $2\times 10^{-3}$ and significantly differs from zero, whereas the value of $|\delta|$ is of order $10^{-4}$ and is consistent with zero within one standard deviation \cite{pdg}.
We neglect $\delta$ in further calculations.
As shown in Sect.\ \ref{II.B}, corrections due to a possible slight difference of coefficients multiplying $|K_S\ra$ and $|K_L\ra$ in Eq. (\ref{eq3}) with nonzero $\delta$ can marginally affect only the non-deformed part of the decay intensity, and we neglect it in what follows.  

In the basis (\ref{eq3}) one gets an expression for $|\psi\ra$
\be
\label{eq4}
|\psi\ra & = & N \left(|K_L(\mathbf p)\ra|K_S(-\mathbf p)\ra - |K_S(\mathbf p)\ra|K_L(-\mathbf p)\ra\right),
\ee
where the normalization factor $N$ depends on $\varepsilon$ and $\delta$.

The decay rate into final states $|f_{1(2)}\ra$ at times $t_{1(2)}$ is equal to the squared transition amplitude of $|\psi\ra$ to the final state of both kaons
\begin{align*}
    I(f_1,t_1;f_2,t_2)=|\la f_1,t_1;f_2,t_2|H|\psi\ra|^2
\end{align*}
Hereon we consider the case of identical final states $|\pi^+\pi^-\ra$ for the $K_S$ and $K_L$ decays, which is advantageous for a possible detection of novel, $\cal CPT$-violating effects, due to the $\cal CP$-suppression of the $K_L\rightarrow \pi^+\pi^-$ decays.
Since we consider only decays to the specific final state $\pi^+\pi^-$, the decay widths $\Gamma_{S,L}$ refer to decays of $K_{S,L}$ to this final state.

Writing the kaons states in terms of plane waves, after changing variables into $T=t_1+t_2$ and $\Delta t = |t_1-t_2|$, and integrating over T from $\Delta t$ to infinity, one gets
\be
\label{eq4.1}
I(\Delta t) = \frac{4|f_L|^2|f_S|^2}{\overline\Gamma}\left[e^{-\Gamma_L\Delta t}+e^{-\Gamma_S\Delta t}-2e^{-\overline\Gamma\Delta t}\cos(\Delta E\Delta t)\right],
\ee
where
\be
f_{S,L} & = & \la\pi^+\pi^-|H|K_{S,L}\ra
\label{eq11}
\ee
and $\Delta E=E_L-E_S$, $\overline\Gamma=\frac{1}{2}(\Gamma_L+\Gamma_S)$.
The identity of the final states of both kaon decays ensures that there is no additional phase in the oscillation term in Eq. (\ref{eq4.1}).

The ratio of the amplitudes in Eq. (\ref{eq11}) is usually denoted by $\eta_{+-}=f_L/f_S=|\eta_{+-}|e^{i\varphi}$ and is equal to the sum of two parameters quantifying two mechanisms of $\cal CP$ violation.
The indirect, measured by $\varepsilon$, is due to mixing of $K^0$ and $\overline K^0$, and the direct one, measured by $\varepsilon^\prime$, is due to decays of $K_S$ and $K_L$ to states with $CP=-1$ and $CP=+1$, respectively, such that $\eta_{+-}=\varepsilon+\varepsilon^\prime$.
Direct $\cal CP$ violation is much weaker, such that $\varepsilon^\prime$ is of order $10^{-6}$, but is significantly different than zero \cite{pdg}.
In further considerations we omit $\varepsilon^\prime$, for simplicity.

\subsection{Entangled pairs of neutral kaons in  $\omega$-model}

As it was argued by authors of Ref. \cite{bernabeu_2}, Lorentz and $\cal CPT$ invariance violation symmetry of quantum gravitational origin may lead to correction to (\ref{eq1}).
Such correction is parameterized by a complex parameter $\omega$ and the modified state reads 
\be
\label{eq2}
|\psi\ra_\omega & = & \frac{1}{\sqrt 2}\left(|K^0(\mathbf p)\ra|\overline K^0(-\mathbf p)\ra - |\overline K^0(\mathbf p)\ra|K^0(-\mathbf p)\ra \right) \nn \\
         & + & \frac{\omega}{\sqrt 2}\left(|K^0(\mathbf p)\ra|\overline K^0(-\mathbf p)\ra + |\overline K^0(\mathbf p)\ra|K^0(-\mathbf p)\ra\right).
\ee
The complex parameter $\omega$ accounts for a possible $\cal CPT$ violation and its magnitude may be of the order $|\omega|\simeq \sqrt{(m_K^2/\kappa)/\Delta\Gamma}$, where $\Delta\Gamma=\Gamma_S-\Gamma_L$ and $\kappa$ is the LIV parameter, presumably of order of Planck mass.
Recent experimental value of $|\omega|$ is of order $10^{-4}$ and is consistent with zero \cite{kloe2}.

Expressing the state $|\psi\ra$ in the basis of the neutral kaon mass eigenstates $|K_S\ra$ and $|K_L\ra$ one gets
\be
\label{eq4.2}
|\psi\ra_\omega & = & N\Bigg[\left(|K_L(\mathbf p)\ra|K_S(-\mathbf p)\ra - |K_S(\mathbf p)\ra|K_L(-\mathbf p)\ra\right) \nn \\
         & + & \omega\left(|K_S(\mathbf p)\ra|K_S(-\mathbf p)\ra - |K_L(\mathbf p)\ra|K_L(-\mathbf p)\ra\right)\Bigg],
\ee
Using the same procedure for calculating decay intensity as in the standard case, modified intensity $I_\omega(\Delta t)$.
After developing our formalism, in Sec. IV.B we give a comparison of the two-kaon decay intensity of Ref. \cite{bernabeu_2} and the one derived in the next Section below.  

\section{Entangled pairs of neutral kaons in $\kappa$-deformed field theory}

In this section we will consider theories with deformed $\cal CPT$ symmetry, with deviations from the standard $\cal CPT$-invariance arising in the case of moving particles. Specifically, in what follows we will use the theory with $\kappa$-deformation derived and discussed in details in \cite{Arzano:2019toz} and
\cite{bevilacqua_1}, but the formalism we are using can be easily extended to more general class of deformed theories, in which one has to do with non-linearly modified momentum composition and the law of momentum conservation.

\subsection{Formalism}


A two-particle state consisting of a particle and antiparticle can be constructed from the first principles in the framework of the $\kappa$-deformed theory of scalar fields \cite{bevilacqua_1}.

If the particle's four-momentum amounts to $p=(E,\mathbf p)$ then for the antiparticle it is given by its antipodal map $-S(p)= (-S_0(p), -\mathbf S( p))$, where
\be
\label{eq5}
S_0(p) & = & -p_4+\frac{\kappa^2}{E+p_4} \nn \\
     & \simeq & -E+\frac{\mathbf p^{\,2}}{\kappa}+{\cal O}(1/\kappa^2) \nn \\
\mathbf S( p) & = & -\frac{\kappa\mathbf p}{E+p_4} \nn \\
          & \simeq & -\mathbf p+\frac{E\mathbf p}{\kappa} +{\cal O}(1/\kappa^2),
\ee
where $p_4=\sqrt{\kappa^2+m^2}$.

Now let us consider two kaons produced in decay of a $\phi_0(1020)$ resonance, in its centre-of-mass frame.
In the deformed theory instead of the state $|\psi\ra$ (\ref{eq1}) we have its deformed counterpart
\be
\label{eq6}
|\psi\ra_\kappa & = & \frac{1}{\sqrt{2}}\left(|K^0(\mathbf p)\ra|\overline K^0(\mathbf S( p))\ra-|\overline K^0(-\mathbf S(p))\ra|K^0(-\mathbf p)\ra\right) \nn \\
         & = & 2^{-3/2}
         \Big(
         |K_L(-\mathbf S(p))\ra|K_L(-\mathbf p)\ra - |K_L(\mathbf p)\ra|K_L(\mathbf S(p))\ra
         \nn \\
         &  &
         +|K_L(-\mathbf S(p))\ra|K_S(-\mathbf p)\ra - |K_S(\mathbf p)\ra|K_L(\mathbf S(p))\ra
         \nn \\
         &  &
         -|K_S(-\mathbf S(p))\ra K_L(-\mathbf p)\ra +
          |K_L(\mathbf p)\ra|K_S(\mathbf S(p))\ra
         \nn \\
         & &
        -|K_S(-\mathbf S(p))\ra|K_S(-\mathbf p)\ra + |K_S(\mathbf p)\ra|K_S(\mathbf S(p)\Big).
        \ee
Notice that, like in the non-deformed case, the total spatial momentum of the two particles is zero since $\mathbf{p} \oplus \mathbf S({p}) = \mathbf S({p})\oplus \mathbf{p} = 0$. The energy is however always positive, so that particles with spatial momentum $\mathbf{p}$ or $-\mathbf{p}$ have energy $E$, while particles with momentum $\mathbf S({p})$ or $-\mathbf S(p)$ have energy $-S_0(p)$ (which is positive on-shell). 
Notice also that in the mass-eigenstate representation the terms proportional to $K_L\,K_L$ and $K_S\,K_S$ appear in the deformed state (\ref{eq6}).

We now boost the two-kaon state \eqref{eq6} with respect to the centre-of-mass frame by large boost with parameter $\gamma\gg1$. 
We boost the two-particle state as described in \cite{Arzano:2022ewc}, \cite{bevilacqua_2}, considering only terms up to first order in the $1/\kappa$ expansion. For simplicity we consider the limit of kaons produced at rest in the centre-of-mass frame. This is justified since $\frac{|\mathbf{p}|}{m} \sim 0.01$. After the boost, both kaons are travelling in the same direction (although with different modulus of the spatial momentum).

In what follows all energies, widths and decay times refer to this boosted frame. Explicit formulas for boosted quantities are given at the end of this subsection.
The boosted state can be written as
\be
\label{boostedstate}
|\psi\ra_\kappa
         & = & 2^{-3/2}
         \Big(
         |K_L(-\mathbf S(p))\ra|K_L(\widehat{\mathbf p})\ra - |K_L(\mathbf p)\ra|K_L(-\widehat{\mathbf S(p)})\ra
         \nn \\
         &  &
         +|K_L(-\mathbf S(p))\ra|K_S(\widehat{\mathbf p})\ra - |K_S(\mathbf p)\ra|K_L(-\widehat{\mathbf S(p)})\ra
         \nn \\
         &  &
         -|K_S(-\mathbf S(p))\ra K_L(\widehat{\mathbf p})\ra +
          |K_L(\mathbf p)\ra|K_S(-\widehat{\mathbf S(p)})\ra
         \nn \\
         & &
        -|K_S(-\mathbf S(p))\ra|K_S(\widehat{\mathbf p})\ra + |K_S(\mathbf p)\ra|K_S(-\widehat{\mathbf S(p)}\Big).
        \ee
where the $\,\, \widehat{} \,\,$ over the spatial momenta on the rightmost vector in each tensor product is just a notation to remind that the deformed boost has been used \cite{Arzano:2022ewc}, \cite{bevilacqua_2}. The spatial momenta in the arguments are the physical momenta.

Let us now consider the decays of kaons to the final states $|f_1\ra$ and $|f_2\ra$ in times $t_1$ and $t_2$, respectively, and write down all partial amplitudes contributing to the overall decay amplitude $\la f_1,t_1; f_2,t_2|H|\psi\ra_\kappa$
\be
\la f_1,t_1; f_2,t_2|H|\psi\ra_\kappa & = & Q_1-Q_2+Q_3-Q_4-Q_5+Q_6-Q_7+Q_8,
\label{eq7}
\ee
where
\be
Q_1 & = & \la f_1,t_1; f_2,t_2|H| K_L(-\mathbf S(p)),K_L(\widehat{\mathbf p})\ra = \la f_1|K_L\ra\la f_2|K_L\ra e^{-iE_Lt_1-\Gamma_Lt_1+i\zeta t_1}e^{-iE_Lt_2-\Gamma_Lt_2} \nn \\
Q_2 & = & \la f_1,t_1; f_2,t_2|H| K_L(\mathbf p),K_L(-\widehat{\mathbf S(p)})\ra = \la f_1|K_L\ra\la f_2|K_L\ra e^{-iE_Lt_1-\Gamma_Lt_1}e^{-iE_Lt_2-\Gamma_Lt_2+i\zeta t_2} \nn \\
Q_3 & = & \la f_1,t_1; f_2,t_2|H| K_L(-\mathbf S(p)),K_S(\widehat{\mathbf p})\ra = \la f_1|K_L\ra\la f_2|K_S\ra e^{-iE_Lt_1-\Gamma_Lt_1+i\zeta t_1}e^{-iE_St_2-\Gamma_St_2} \nn \\
Q_4 & = & \la f_1,t_1; f_2,t_2|H| K_S(\mathbf p),K_L(-\widehat{\mathbf S(p)})\ra = \la f_1|K_S\ra\la f_2|K_L\ra e^{-iE_St_1-\Gamma_St_1}e^{-iE_Lt_2-\Gamma_Lt_2+i\zeta t_2} \nn \\
Q_5 & = & \la f_1,t_1; f_2,t_2|H| K_S(-\mathbf S(p)),K_L(\widehat{\mathbf p})\ra = \la f_1|K_S\ra\la f_2|K_L\ra e^{-iE_St_1-\Gamma_St_1+i\zeta t_1}e^{-iE_Lt_2-\Gamma_Lt_2} \nn \\
Q_6 & = & \la f_1,t_1; f_2,t_2|H| K_L(\mathbf p),K_S(-\widehat{\mathbf S(p)})\ra = \la f_1|K_L\ra\la f_2|K_S\ra e^{-iE_Lt_1-\Gamma_Lt_1}e^{-iE_St_2-\Gamma_St_2+i\zeta t_2} \nn \\
Q_7 & = & \la f_1,t_1; f_2,t_2|H| K_S(-\mathbf S(p)),K_S(\widehat{\mathbf p})\ra = \la f_1|K_S\ra\la f_2|K_S\ra e^{-iE_St_1-\Gamma_St_1+i\zeta t_1}e^{-iE_St_2-\Gamma_St_2} \nn \\
Q_8 & = & \la f_1,t_1; f_2,t_2|H| K_S(\mathbf p),K_S(-\widehat{\mathbf S(p)})\ra = \la f_1|K_S\ra\la f_2|K_S\ra e^{-iE_St_1-\Gamma_St_1}e^{-iE_St_2-\Gamma_St_2+i\zeta t_2}.
\label{eq8}
\ee

In Eq. (\ref{eq8}), $E_{L(S)}$ stand for boosted energies and $\Gamma_{L(S)}$ for decay widths of the $K_{L(S)}$, respectively.
The quantity $\zeta=\mathbf p^{\,2}/\kappa$ accounts for the order $1/\kappa$ correction to energy due to deformation. Boosting a two-particle system in the deformed case is, in general, quite nontrivial \cite{Arzano:2022ewc}, \cite{bevilacqua_2}, so that different $\zeta_i$ would be needed, one for each $Q_i$. In this case, however, we are only interested in the first order contribution in $1/\kappa$, and one can show that the effect of deformation can be fully captured by the single parameter $\zeta$. Indeed, the different $\zeta_i$ differ by multiples of $m_L \Delta m/\kappa$, which can be safely ignored.

The two-kaon decay amplitude (\ref{eq7}) contains terms with the correct $\cal CPT$ (we denote it by $C$) and wrong $\cal CPT$ parity (denoted by $W$).
The first consists of mixed terms $|K_S\ra|K_L\ra$ and $|K_L\ra|K_S\ra$, and the second one of terms $|K_L\ra|K_L\ra$ and $|K_S\ra|K_S\ra$:
\be
C & = & Q_3-Q_4-Q_5+Q_6 \nn \\
W & = & Q_1-Q_2-Q_7+Q_8
\label{eq9}
\ee
The decay intensity $I=|\la f_1,t_1;f_2,t_2|H|\psi\ra_\kappa|^2$ can be then neatly expressed as intensities of the correct-$\cal CPT$, wrong-$\cal CPT$, and interference between the two.
After changing variables into $T=t_1+t_2$ and $\Delta t=|t_1-t_2|$, and integrating over $T$ from $\Delta t$ to infinity, one gets
\be
I_\kappa(\Delta t)=|C(\Delta t)|^2+|W(\Delta t)|^2+2\Re [C(\Delta t) W(\Delta t)^\ast].
\label{eq10}
\ee

In our further calculations we consider the case of the same final states $|\pi^+\pi^-\ra$ of the $K_S$ and $K_L$ decays.
Formulas for other combinations of final states can be derived in a similar way.

Explicit expressions for all terms of Eq. (\ref{eq10}) read:
\be
|C(\Delta t)|^2 & = & \frac{2}{\overline\Gamma}|f_L|^2|f_S|^2\big(1+\cos(\zeta\Delta t)\big)\left[e^{-\Gamma_L\Delta t}+e^{-\Gamma_S\Delta t}-2e^{-\overline\Gamma\Delta t}\cos(\Delta E\Delta t)\right],
\label{eq12}
\ee
\be
|W(\Delta t)|^2 & = & 2\big(1-\cos(\zeta \Delta t)\big)\Bigg[\frac{|f_L|^4}{\Gamma_L}e^{-\Gamma_L\Delta t}+\frac{|f_S|^4}{\Gamma_S}e^{-\Gamma_S\Delta t} \nn \\
  & - & 2\Re\,(f_L^2\bar f_S^2)\frac{e^{-\overline\Gamma\Delta t}}{\overline\Gamma^2+(\Delta E)^2}\left(\Delta E\sin(\Delta E\Delta t)-\overline\Gamma\cos(\Delta E\Delta t)\right) \Bigg],
\label{eq13}
\ee
and
\begin{align}
    &2\Re\,[C(\Delta t)W^\ast(\Delta t)]  =  -16\sin(\zeta\Delta t)\Bigg[|f_L|^2\Im\,(\bar f_Lf_S)\frac{\Delta E}{(\Gamma_L+\overline\Gamma)^2+(\Delta E)^2}e^{-\Gamma_L\Delta t} \nn \\
  & +  |f_S|^2\Im\,(f_L\bar f_S)\frac{\Delta E}{(\Gamma_S+\overline\Gamma)^2+(\Delta E)^2}e^{-\Gamma_S\Delta t} \nn \\
  & +  |f_L|^2\Im\,(\bar f_Lf_S)\frac{e^{-\overline\Gamma\Delta t}}{(\Gamma_L+\overline\Gamma)^2+(\Delta E)^2}\left[(\Gamma_L+\overline\Gamma)\sin(\Delta E\Delta t)+\Delta E\cos(\Delta E\Delta t)\right] \nn \\
  & +  |f_S|^2\Im\,(f_L\bar f_S)\frac{e^{-\overline\Gamma\Delta t}}{(\Gamma_S+\overline\Gamma)^2+(\Delta E)^2}\left[(\Gamma_S+\overline\Gamma)\sin(\Delta E\Delta t)+\Delta E\cos(\Delta E\Delta t)\right] \Bigg], 
\label{eq14}
\end{align}

Since the decay $K_L\rightarrow\pi^+\pi^-$ violates $\cal CP$, its amplitude is suppressed with respect to $K_S\rightarrow\pi^+\pi^-$ by a factor $|\varepsilon|=|f_L/f_S|\simeq 2\times 10^{-3}$.

\subsection{Boosted quantities}

From the phenomenological point of view, an important feature of the $\kappa$-deformation is the dependence of the deformation-induced oscillation frequency $\zeta$ on energy.
Since $\zeta=\mathbf p^{\,2}/\kappa$, any effects of deformation vanish in the particle's rest frame. 
Let us thus write all the boosted quantities explicitly in terms of $\gamma$.
As shown in Ref. \cite{bevilacqua_2}, for strong boosts and neglecting tiny effects of non-zero transverse momenta, the arguments of the oscillating terms in Eqs. (\ref{eq12}-\ref{eq14}) transform as
\be
(\Delta E\Delta t)^\prime & = & \gamma^2\,\Delta m\Delta t \nn \\
(\zeta \Delta t)^\prime & = & \gamma^3m^2\Delta t/\kappa.
\label{eq15}
\ee
Decay widths contract as $\Gamma^\prime=\Gamma/\gamma$, decay times dilate as $t^\prime=\gamma t$ and energies increase as $E^\prime=\gamma m$.
Therefore the arguments $\Gamma\Delta t$ of exponentials in Eqs. (\ref{eq12}-\ref{eq14}) are Lorentz-invariant.

\section{Observability of deformations}\label{II.B}

Having obtained explicit expressions for deformed decay intensity, let us now turn to the question as to if the effect of deformation could be observable.

\subsection{Dominating terms and order of magnitude considerations}

Let us start with Eqs.\ (\ref{eq12}-\ref{eq14}), which can be rewritten as follows:
\be
|C(\Delta t)|^2 & = & \frac{2}{\overline\Gamma}|f_S|^4|\eta_{+-}|^2\big(1+\cos(\zeta\Delta t)\big)\left[e^{-\Gamma_L\Delta t}+e^{-\Gamma_S\Delta t}-2e^{-\overline\Gamma\Delta t}\cos(\Delta E\Delta t)\right],
\label{eq13p}
\ee
\be
|W(\Delta t)|^2 & = & 2|f_S|^4 \big(1-\cos(\zeta \Delta t)\big)\Bigg[\frac{|\eta_{+-}|^4}{\Gamma_L}e^{-\Gamma_L\Delta t}+\frac{1}{\Gamma_S}e^{-\Gamma_S\Delta t} \nn \\
  & - & 2|\eta_{+-}|^2\frac{e^{-\overline\Gamma\Delta t}}{\overline\Gamma^2+(\Delta E)^2}\big(\Delta E\sin(\Delta E\Delta t)-\overline\Gamma\cos(\Delta E\Delta t)\big) \Bigg],
\label{eq14p}
\ee
and
\be
2\Re\,[C(\Delta t)W^\ast(\Delta t)] & = & -16|f_S|^4\sin(\zeta\Delta t)\Bigg[|\eta_{+-}|^3\frac{\Delta E}{(\Gamma_L+\overline\Gamma)^2+(\Delta E)^2}e^{-\Gamma_L\Delta t} \nn \\
  & + & |\eta_{+-}|\frac{\Delta E}{(\Gamma_S+\overline\Gamma)^2+(\Delta E)^2}e^{-\Gamma_S\Delta t} \nn \\
  & + & |\eta_{+-}|^3\frac{e^{-\overline\Gamma\Delta t}}{(\Gamma_L+\overline\Gamma)^2+(\Delta E)^2}\left[(\Gamma_L+\overline\Gamma)\sin(\Delta E\Delta t+\varphi)+\Delta E\cos(\Delta E\Delta t+\varphi)\right] \nn \\
  & + & |\eta_{+-}|\frac{e^{-\overline\Gamma\Delta t}}{(\Gamma_S+\overline\Gamma)^2+(\Delta E)^2}\left[(\Gamma_S+\overline\Gamma)\sin(\Delta E\Delta t+\varphi)+\Delta E\cos(\Delta E\Delta t+\varphi)\right] \Bigg]. \nn \\
\quad
\label{eq15p}
\ee
In the correct-parity term (\ref{eq13p}) one finds a common factor $|\eta_{+-}|^2$ multiplying the whole expression, because in the initial state there is always one short-lived $K_S$ and one long-lived $K_L$, both decaying into the $\pi^+\pi^-$ final states, where the first is $\cal CP$-allowed and the second one is $\cal CP$-suppressed.
The wrong-parity term (\ref{eq14p}) consists of either two $K_S$, whose decays are $\cal CP$-allowed, or two $K_L$ decaying in the $\cal CP$-suppressed mode.
The latter ones thus incorporate the $|\eta_{+-}|^2$ factors in the amplitudes ($|\eta_{+-}|^4$ in decay widths) and the former ones are entirely favoured and thus have no $\eta_{+-}$-dependent coefficients.
The interference term (\ref{eq15p}) contains coefficients proportional either to $|\eta_{+-}|$ for interference of $|K_L\ra|K_L\ra$ with $|K_L\ra|K_S\ra$ states, or $|\eta_{+-}|^3$ for interference of $|K_S\ra|K_S\ra$ with $|K_L\ra|K_S\ra$.

Inspecting expressions (\ref{eq13p} - \ref{eq15p}) one can see that in the kaon rest frames, where $\zeta=0$, the terms of wrong parity $|W(\Delta t)|^2$, and interference between the correct and wrong parity terms $\Re[C(\Delta t)W^\ast(\Delta t)]$ disappear and reproduce the undeformed case.

Further simplification, suitable for subsequent quantitative assessments, can be done by neglecting terms proportional to $1/\gamma$ and keeping only those of order $\sim\gamma$.
One thus gets an approximation valid for large boosts $\gamma$

\be
|C(\Delta t)|^2 & \simeq & \frac{|\eta_{+-}|^2|f_S|^4}{\overline\Gamma} 2\big(1+\cos(\zeta\Delta t)\big)\left[e^{-\Gamma_L\Delta t}+e^{-\Gamma_S\Delta t}-2e^{-\overline\Gamma\Delta t}\cos(\Delta E\Delta t)\right],
\label{eq13pp}
\ee
\be
|W(\Delta t)|^2 & \simeq & \frac{|\eta_{+-}|^2|f_S|^4}{\overline\Gamma} \big(1-\cos(\zeta\Delta t)\big)\left[|\eta_{+-}|^2\, e^{-\Gamma_L\Delta t}+\frac{1}{|\eta_{+-}|^2}\,e^{-\Gamma_S\Delta t}\right].
\label{eq14pp}
\ee
The interference term is negligible in this approximation.

In Eqs. (\ref{eq13pp},\ref{eq14pp}) a common factor $|\eta_{+-}|^2|f_S|^4/\overline\Gamma$ is put in front of the time-dependent expressions, in order to make both terms directly comparable.
It can be seen that $|W(\Delta t)|^2$ is strongly dominated by term proportional to $e^{-\Gamma_S\Delta t}$, for reasons explained above, thus making it orders of magnitude larger than this for $|C(\Delta t)|^2$.

\subsection{Relation to the $\omega$-model}

The decay intensity in Eq. (8) of Ref. \cite{bernabeu_2} contains three terms
\begin{align}
\label{beq1}
    I_\omega(\Delta t)=\left|\left\langle\pi^{+} \pi^{-} \mid K_S\right\rangle\right|^4|C|^2\left|\tilde{\eta}_{+-}\right|^2\left[I_1+I_2+I_{12}\right],
\end{align}
where
\begin{align}
\label{beq2}
    I_1(\Delta t)&=\frac{e^{-\Gamma_S \Delta t}+e^{-\Gamma_L \Delta t}-2 e^{-\left(\Gamma_S+\Gamma_L\right) \Delta t / 2} \cos (\Delta M \Delta t)}{\Gamma_L+\Gamma_S} \nonumber \\
    I_2(\Delta t)&=\frac{|\omega|^2}{\left|\tilde{\eta}_{+-}\right|^2} \frac{e^{-\Gamma_S \Delta t}}{2 \Gamma_S}
\end{align}
and
\begin{align}
\label{beq3}
I_{12}(\Delta t)&=  -\frac{4}{4(\Delta M)^2+\left(3 \Gamma_S+\Gamma_L\right)^2} \frac{|\omega|}{\left|\tilde{\eta}_{+-}\right|}\nonumber \\
& \times\left\{2 \Delta M\left[e^{-\Gamma_S \Delta t} \sin \left(\phi_{+-}-\Omega\right)-e^{-\left(\Gamma_S+\Gamma_L\right) \Delta t / 2} \sin \left(\phi_{+-}-\Omega+\Delta M \Delta t\right)\right]\right.\nonumber \\
& \left.-\left(3 \Gamma_S+\Gamma_L\right)\left[e^{-\Gamma_S \Delta t} \cos \left(\phi_{+-}-\Omega\right)-e^{-\left(\Gamma_S+\Gamma_L\right) \Delta t / 2} \cos \left(\phi_{+-}-\Omega+\Delta M \Delta t\right)\right]\right\}.
\end{align}
In Eqs. (\ref{beq1}-\ref{beq3}) the original notation is kept where $\tilde{\eta}_{+-}=1/\eta_{+-}$ and $\Delta M=-\Delta m$, $\Omega$ stands for the phase of $\omega$, and $\phi_{+-}=-\varphi$ for phase of $\tilde{\eta}_{+-}$.

In Eq. (\ref{beq2}, $I_1(\Delta t)$ corresponds to the correct-parity term $|C(\Delta t)|^2$ (\ref{eq13p}) but does not contain the deforming factor $(1+\cos(\zeta\Delta t))$ which makes an important difference between these two models. 
The $I_2(\Delta t)$ consists of a single term proportional to $e^{-\Gamma_S}$ whereas in the wrong-parity term $|W(\Delta t)|^2$ (\ref{eq14p}) there are two additional terms where the one proportional to $e^{-\Gamma_L}$ is strongly suppressed by $|\eta_{+-}|^2$ factor. 
The deforming term $1-\cos(\zeta\Delta t)$ (\ref{eq14p}) modifies the expression analogously to $\omega^2$ (\ref{beq2}).
Finally, the interference $I_{12}(\Delta t)$ (\ref{beq3}) corresponds to $2\Re[C(\Delta t) W^\ast(\Delta t)]$ (\ref{eq15p}).
Their structures are similar but the first depends linearly on $\omega$ and the second one on $\sin(\zeta\Delta t)$.

Apart from all formal similarities, the fundamental physical difference is that in the model presented here the $\cal CPT$ violation is present only in the moving frame and it disappears in the kaon rest frame where $|\psi\ra_\kappa \rightarrow |\psi(\mathbf p=\mathbf 0)\ra$ for $\mathbf p\rightarrow \mathbf 0$, cf. Eqs. (\ref{eq1}, \ref{eq6}).
The $\omega$-model of Ref. \cite{bernabeu_2} is formulated for kaons at rest where the violation of $\cal CPT$ is allowed.

\subsection{Numerical estimates of deformation}

Deformed oscillations in the decay-intensity spectrum provide new opportunities to estimate the deformation scale $\kappa$. 
In particular, both intensities (\ref{eq13pp}) and (\ref{eq14pp}) hint at what an experimentalist should look at to find signatures of deformation in the decay spectra of interfering kaons.
In the following, let us discuss two semi-quantitative approaches to the problem, bearing in mind that complete qualitative analyses can only be done using simulations for specific experimental setups. 

\subsubsection{Frequency of the deformed oscillations}

The deformation-dependent oscillations, governed by $1\pm\cos(\zeta\Delta t)$ factors in $|C(\Delta t)|^2$ and $|W(\Delta t)|^2$ terms, have not yet been attempted to be observed experimentally.
We are not aware of any observations made with boosts high enough to make such effects potentially visible.
Let us formulate the necessary, order-of-magnitude conditions for such an observation without any sophisticated analysis of the shape of the decay time intensity.

Considering an experiment with a large boost one may expect to see an effect, provided that the deformed oscillations with frequency $\zeta$ are not too slow compared to the undeformed ones with the frequency $\Delta E$.
According to Eq. (\ref{eq15}), boosted periods of normal oscillations, amounting to $1/\Delta E$, extend from $10^{16}$ GeV$^{-1}$ down to $10^{10}$ GeV$^{-1}$ for the corresponding Lorentz $\gamma$ in the range from $10^2$ to $10^5$.
For the same range of $\gamma$, the boosted deformed oscillation periods $1/\zeta$ extend from $10^{13}$ GeV$^{-1}$ down to $10^{10}$ GeV$^{-1}$.
It is thus seen that the two oscillation periods, $1/\Delta E$ and $1/\zeta$, become comparable only for large $\gamma\sim 10^{5}$, corresponding to protons of an energy of $100$ TeV, yet unattainable in accelerators.

The unobservability of deformed oscillations can provide us an upper limit for $\kappa$.
If one requires that $\cos(\zeta\Delta t)$ varies very slowly compared to $\cos(\Delta E\Delta t)$, which means $\zeta \ll \Delta E$, by using Eq. (\ref{eq15}) one translates this inequality into condition
\be
\kappa \gg \gamma\times 0.5\times 10^{14}\; \mbox{GeV}.
\label{eq16}
\ee
For example, using energy of $10$ TeV one gets a lower limit of at most $10^{17}$ GeV, still located below the Planck energy scale $10^{19}$ GeV.
This provides us with an exciting possibility to limit $\kappa$ experimentally by pushing further up future accelerating technologies.

\subsubsection{Intensity of the wrong-parity oscillations}

Another way of limiting the value of deformation parameter $\kappa$ may be to exploit strong amplification of the wrong-parity term $|W(\Delta t)|^2$ due to the double $\cal CP$-favoured decay of the $|K_S\ra|K_S\ra$ state where each $K_S$ decays into the $\pi^+\pi^-$ pair.
Since $|\eta_{+-}|^2=0.4\times 10^{-5}$, one may expect that unbservability of the wrong-parity term requires $1-\cos(\zeta\Delta t)$ to be limited from above by $\sigma_W < 10^{-6}$, depending on available experimental accuracy.
Here we assume that the suppression factor $\sigma_W$ is an order of magnitude smaller than the amplification $|\eta_{+-}|^2$ in the wrong-parity term (\ref{eq14pp}). 
Realistic values of $\sigma_W$ depend on many factors, such as the statistical errors and overall shape of the intensity spectrum due to the apparatus acceptance.

 The requirement 
 \be
 1-\cos(\zeta\Delta t) < \sigma_W
 \label{eq17n}
 \ee
translates into a condition
\be
\kappa>\frac{\gamma^3m^2\Delta t_r}{2\sqrt{\sigma_W}},
\label{eq17}
\ee
where $t_r$ refers to decay time in the kaon rest frame.
For $\gamma\sim 10^3$, $\sigma_W=10^{-6}$ and $\Delta t_r=\tau_S=1.4\times 10^{14}$ GeV$^{-1}$, the condition (\ref{eq17}) provides with a limitation $\kappa > 0.2\times 10^{25}\; \mbox{GeV}$,
well above the Planck scale.
Interestingly, this limit may be much tightened from below by using smaller boosts.
It can be also lowered by using smaller decay times, depending on the experimental time resolution.
For currently available time resolutions of $\sigma_t=4.5$ fs \cite{lhcb_res} the lower limit on $\kappa$ can be calculated from Eq. (\ref{eq17}) by taking $\gamma\Delta t_r=\sigma_t$ which gives 
\be
\kappa > 0.2\times 10^{18}\; \mbox{GeV}.
\label{eq18}
\ee
Strong amplification of $|W(\Delta t)|^2$ (\ref{eq14pp}) makes this method potentially more efficient in detecting the wrong $\cal CPT$ than the comparison of oscillation frequencies.
The $|\eta_{+-}|^{-2}$ factor multiplies the short-time component $e^{-\Gamma_S\Delta t}$ in Eq. (\ref{eq14pp}) where it decreases fast with $\Delta t$.
This may be also advantageous due to higher sensitivity to shape compared to slowly varying component $e^{-\Gamma_L\Delta t}$.

\subsubsection{Resonant oscillations}

Potentially the most stringent estimation of $\kappa$ is provided at high energy when the deformed frequency $\zeta$ becomes equal to the normal oscillation frequency $\Delta E$.
According to Eq. (\ref{eq15}), this kind of resonance happens at $\gamma^\ast=(\Delta m/m^2)\kappa$.
For instance, if the deformation $\kappa$ is of the order of the Planck's mass, $\kappa\simeq 10^{19}$ GeV, then $\gamma^\ast\simeq 0.2\times 10^{5}$, corresponding to energies $10^2 - 10^3$ TeV.
Using condition $\zeta=\Delta E$ as a prediction for $\kappa$ one gets
\be
\kappa = \gamma\times 0.45\times 10^{14}\; \mbox{GeV}
\label{eq19}
\ee
which represents a point estimate of $\kappa$, and not a lower limit, thus demonstrating the potential power of the method.

\vskip 5mm
Doing predictions for future experiments it has to be realized, however, that the precision of the estimates (\ref{eq16}), (\ref{eq18}) and (\ref{eq19}) predominantly depends on the uncertainty of determination of the energy $E$ which directly propagates to the uncertainty of the Lorentz boost $\gamma=E/m$. 
In current high-energy experiments, the resolution of energy or momentum is typically at the percent level.
Resolutions in $\Delta m$ and $m$ are $2\times 10^{-3}$ and $3\times 10^{-5}$ \cite{pdg}, respectively, and can be neglected in these predictions.
The uncertainty of energy measurement thus critically affects the precision of $\kappa$ and for the energy's uncertainty $10^{-2}$ one only determines our estimates with an error of $\pm \gamma\times 10^{12}$.
To make this method efficient, one has to significantly improve the accuracy of the energy measurement. 
 
\section{Conclusions}

In this paper, we investigated kaon interferometry in the quantum gravity-inspired model with deformed (as opposed to manifestly violated) continuous (Poincar\'e) and discrete ($\cal CPT$) symmetries. We found that in this case the decay intensity of kaon -- anti-kaon pair has three contributing terms: the right-parity, the wrong-parity, and the interference between them, all of which are affected by deformation. The magnitude of deviations from the predictions of the standard, undeformed, $\cal CPT$-invariant theory is proportional to the  momenta carried by kaons, and, if real, can be in principle observed. 

A distinctive feature of the approach of the present paper is its derivation from the concepts of non-commutative geometry and deformed conservation rules. 
Our prediction for the boost-dependent violation of the $\cal CPT$ symmetry was derived from the first principles.
Its dependence on energy and decay times provides us with a clear prescription on how to examine this scenario on the fast-moving technology frontiers of energy and precision. 

The system of kaons offers to possible observational opportunities: one associated with an additional momentum-dependent oscillation frequency and another being a consequence of the emergence of a wrong parity term, that is not allowed by the standard $\cal CPT$-invariance. We estimate that the first method provides the bound $\kappa \gtrsim 10^{14}$ GeV in the case of the boost $\gamma \sim 10^5$ (an order of magnitude higher than that attainable with the present accelerator techniques). The second method $\kappa \gtrsim 10^{18}$ GeV, is already close to the expected value of $\kappa$, of the order of Planck mass.
The third method, potentially the most powerful, can predict the value of $\kappa$ and not its lower limit. 
For $\gamma\sim 10^5$ one gets $\kappa \sim 10^{19}$ GeV, so the Planck's energy scale.
This suggests that it is feasible that the theory with deformed $\cal CPT$-invariance might be ruled out, or confirmed in the not-too-distant accelerator experiments \cite{fcc}.
However, to make these measurements conclusive, one has to significantly improve the accuracy of the experimental determination of particles' boost.

Finally, let us compare our model with the one proposed in Ref. \cite{bernabeu_2}. In the parametrization of that paper, the $\omega$ parameter, quantifying an admixture of states with an ill-defined $\cal CPT$, is found only in terms originating from the $|K_L\ra|K_L\ra$ and $|K_S\ra|K_S\ra$ states and thus, apart from the overall normalization, modifies only the wrong-parity and interference terms. 
In our case, the $\kappa$-dependent factors influence oscillations of all terms (\ref{eq13p})-(\ref{eq15p}), each in a different manner.
If identified directly with the factor multiplying the wrong-parity term of the intensity (\ref{eq14p}), the $|\omega|^2\sim 1-\cos (\zeta\Delta t)$ and is energy- and time-dependent.

\section*{Acknowledgment}
For AB and JKG  this work was supported by funds provided by the National Science Center, project no. 2019/33/B/ST2/00050.

\end{document}